\documentclass[twocolumn,showpacs,preprintnumbers,amsmath,amssymb]{revtex4}

\usepackage{graphicx}
\usepackage{dcolumn}
\usepackage{bm}

\begin{document}


\title{Progress in Lunar Laser Ranging Tests of Relativistic Gravity}

\author{James G. Williams}
 \email{james.g.williams@jpl.nasa.gov}
\author{Slava G. Turyshev} 
 \email{turyshev@jpl.nasa.gov}
\author{Dale H. Boggs} 
 \email{dale.h.boggs@jpl.nasa.gov}
\affiliation{
Jet Propulsion Laboratory, California Institute of  Technology,\\
Pasadena, CA 91109, USA
}

\date{\today}

\begin{abstract}
Analyses of laser ranges to the Moon provide increasingly stringent limits on any violation of the equivalence principle (EP); they also enable several very accurate tests of relativistic gravity.  These analyses give an EP test of $\Delta (M_G/M_I)_{EP} =(-1.0\pm1.4)\times10^{-13}$. This result yields a strong equivalence principle (SEP) test of $\Delta (M_G/M_I)_{SEP} =(-2.0\pm2.0)\times10^{-13}$. Also, the corresponding SEP violation parameter $\eta$ is $(4.4\pm4.5)\times 10^{-4}$, where $\eta=4\beta-\gamma-3$ and both $\beta$ and $\gamma$ are post-Newtonian parameters. Using the Cassini $\gamma$, the $\eta$ result yields $\beta-1=(1.2\pm1.1)\times 10^{-4}$. The geodetic precession test, expressed as a relative deviation from general relativity, is $K_{gp}=-0.0019\pm0.0064$. The search for a time variation in the gravitational constant results in $\dot G/G=(4\pm9)\times10^{-13}$ yr$^{-1}$; consequently there is no evidence for local ($\sim$1~AU) scale expansion of the solar system.
\end{abstract}

\maketitle

\section{\label{sec:intro}Introduction}

Einstein's general theory of relativity (GR) began its empirical success in 1915 by explaining the anomalous perihelion precession of Mercury's orbit, using no adjustable theoretical parameters.  Shortly thereafter, Eddington's 1919 observations of star lines-of-sight during a solar eclipse confirmed the doubling of the deflection angles predicted by GR as compared to Newtonian and Equivalence Principle (EP) arguments.  Following these beginnings, the general theory of relativity has been verified at ever-higher accuracy. Thus, microwave ranging to the Viking landers on Mars yielded a $\sim$0.2\%  accuracy via the Shapiro time-delay \cite{[47]}.  Spacecraft and planetary radar observations reached an accuracy of $\sim$0.15\% \cite{anderson02}. Lunar Laser Ranging (LLR) has provided verification of GR improving the accuracy to $\sim$0.05\% via precision measurements of the lunar orbit \cite{[62],science94,[39],[63],[2]}. The astrometric observations of the deflection of quasar positions with respect to the Sun performed with Very-Long Baseline Interferometry (VLBI) improved the accuracy of the GR tests to $\sim$0.045\% \cite{[48],Shapiro_SS_etal_2004}.  The  time-delay experiment with the Cassini spacecraft at a solar conjunction  has tested gravity with a remarkable accuracy of $0.0023\%$  \cite{cassini_ber} in measuring the deviation of the parametrized post-Newto\-nian (PPN) parameter $\gamma$ from its GR value of unity. 

The accuracy of the Cassini result opens a new realm for tests of gravity in the solar system, especially those motivated by the progress in scalar-tensor theories of gravity. In particular, the recent work in scalar-tensor extensions of gravity that are consistent with present cosmological models \cite{[12a],DPV02,[15],[12n]} predicts  deviations of the parameter $\gamma$ from the general relativistic value at levels of 10$^{-5}$ to 10$^{-7}$. This prediction motivates new searches for very small deviations of relativistic gravity from GR in the solar system and provides a robust theoretical paradigm and constructive guidance for experiments that would decrease the uncertainty of PPN parameters.  Thus, in addition to experiments which probe the parameter $\gamma$, any experiment extending the accuracy in measuring parameter $\beta$ is also of great interest.  Today LLR,  the continuing legacy of the Apollo program, is well positioned to address this challenge. 

Motivated by the remarkable accuracy of the Cassini test for the PPN parameter $\gamma$ \cite{cassini_ber}, this Letter reports the results of recent LLR analyses using data to April 2004. The focus is on the improvement of accurate LLR gravity experiments, especially the tests of EP, Strong Equivalence Principle (SEP), PPN parameter $\beta$, and $\dot G$.

\section{Fundamental Physics with LLR}

LLR has a history \cite{science94} dating back to the placement of a retroreflector array on the lunar surface by the Apollo 11 astronauts. Additional reflectors were left by the Apollo 14 and Apollo 15 astronauts, and two French-built reflector arrays were placed on the Moon by the Soviet Luna 17 and Luna 21 missions. LLR accurately measures the time of flight for a laser pulse fired from an observatory on the Earth, bounced off of a corner cube retroreflector on the Moon, and returned to the observatory.  For a general review of LLR see Dickey et al. \cite{science94}.  A comprehensive paper on LLR tests of gravitational physics is Williams et al. \cite{[62]}. A recent test of the EP is in Anderson and Williams \cite{[2]} and other gravitational physics tests are in Williams et al. \cite{jim01}.  An overview of the LLR gravitational physics tests is given by Nordtvedt \cite{ken30}.  Reviews of various tests of relativity, including the contribution by LLR, are given in Will \cite{[58]}. 

The LLR measurements contribute to a wide range of scientific investigations \cite{[62],science94,[9],jj02,williams2001}, and are today solely responsible for the production of the lunar ephemeris. On the fundamental physics front, LLR provides the only current solar system means for testing the SEP---the statement that \emph{all} forms of mass and energy contribute equivalent quantities of inertial and gravitational mass. In addition, LLR is capable of measuring the time variation of Newton's gravitational constant, {\it G}, providing the strongest limit available for the variability of this ``constant.'' LLR can also precisely measure the de Sitter precession---effectively a spin-orbit coupling affecting the lunar orbit in the frame co-moving with the Earth-Moon system's motion around the Sun. Finally, current LLR results are consistent with the GR gravitomagnetic effect on the lunar orbit within 0.1\% of the predicted level \cite{[12n],ken30}.  Thus,  the lunar orbit is a unique laboratory for gravitational physics where each term in the relativistic PPN equations of motion is verified to a very high accuracy.

\subsection{Equivalence Principle Tests}

The Equivalence Principle, the exact correspondence of gravitational and inertial masses, is a central assumption of general relativity and a unique feature of gravitation. It is this principle that predicts identical accelerations of compositionally different objects in the same gravitational field, and also allows gravity to be viewed as a geometrical property of spacetime--leading to the general relativistic interpretation of gravitation. EP tests can therefore be viewed in two contexts: tests of the foundations of the standard model of gravity (i.e. general relativity), or as searches for new physics because, as emphasized in \cite{[12a],DPV02,[15],[12n]}, almost all extensions to the standard model of particle physics generically predict new forces that would show up as apparent violations of the EP. 

The weak form of the EP (the WEP) states that the gravitational properties of strong and electro-weak interactions obey the EP. In this case the relevant test-body differences are their fractional nuclear-binding differences, their neutron-to-proton ratios, their atomic charges, etc. GR and other metric theories of gravity assume that the WEP is exact. However, extensions of the standard model of particle physics that contain new macroscopic-range quantum fields predict quantum exchange forces that generically violate the WEP because they couple to generalized ``charges'' rather than to mass/energy as does gravity \cite{[15],DPV02}.  WEP tests can be conducted with laboratory or astronomical bodies because the relevant differences are in the test-body compositions.

In its strong form the EP is extended to cover the gravitational properties resulting from gravitational energy itself. In other words, it is an assumption about the way that gravity begets gravity, i.e. about the non-linear property of gravitation. Although general relativity assumes that the SEP is exact, alternate metric theories of gravity such as those involving scalar fields, and other extensions of gravity theory, typically violate the SEP \cite{[39],[12n]}. For the SEP case, the relevant test body differences are in the fractional contributions to their masses by gravitational self-energy. To facilitate investigation of a possible violation of the SEP, the ratio between gravitational and inertial masses, $M_G/M_I$, is expressed as
\begin{equation}
\left[\frac{M_G}{M_I}\right]_{SEP}=1+\eta\frac{U}{Mc^2},
\end{equation}	  	
\noindent where $U$ is the body's gravitational self-energy $(U<0)$, $Mc^2$ is its total mass-energy, and $\eta$  is a dimensionless constant \cite{[39]}. Because gravitational self-energy is proportional to $M^2$ (i.e. $U/Mc^2\propto M$) and gravity is so extremely weak, SEP test bodies that differ significantly must have astronomical sizes. Currently, the Earth-Moon-Sun system provides the best arena for testing the SEP with LLR being the only solar system technique available to enable the tests. 

The quasi-Newtonian acceleration of the Moon $(m)$ with respect
to the Earth $(e)$, ${\bf a}={\bf a}_m -  {\bf a}_e$, for the three-body Earth-Moon-Sun (s) system is
\begin{equation}
{\bf a} = - \mu^* {   {\bf r}_{em} \over r^3_{em}} -
\Bigl ({M_G \over M_I} \Bigl )_{\hskip-2pt e}\mu_s {{\bf r}_{es} \over r_{es}^3} +
 \Bigl ({M_G \over M_I} \Bigl )_{\hskip-2pt m}
\mu_s { {\bf r}_{ms} \over r_{ms}^3},  \\\label{eq:range1_m}
\end{equation}

\noindent where  $\mu^*\equiv \mu_e ({M_G /M_I})_m+ \mu_m({M_G /M_I})_e$ and $\mu_k \equiv G M_k$.
The last two terms of Eq.~(\ref{eq:range1_m}) represent the solar effect on the motion of the Moon with respect to the Earth. A violation of the EP would produce a lunar orbit perturbation proportional to the difference in the two $M_G/M_I$ ratios. An SEP test would involve  \cite{[62]}
{}
\begin{eqnarray}
\left[\left(\frac{M_G}{M_I}\right)_e-
\left(\frac{M_G}{M_I}\right)_m\right]_{SEP} &=& \left[\frac{U_e}{M_ec^2}-\frac{U_m}{M_mc^2}\right] \eta \nonumber\\
&=& -4.45\times10^{-10}\eta.
\label{eq:sep_llr}
\end{eqnarray}
In general, $\eta$  is a linear function of seven of the ten PPN parameters, but considering only $\beta$  and  $\gamma$
\begin{equation}
\eta=4\beta-\gamma-3.
\label{3}
\end{equation}
In general relativity $\eta  = 0$. A non-zero value for $\eta$  would produce a displacement of the lunar orbit about the Earth \cite{[39],[58]}; a unit value would cause a 13 meter monthly range modulation \cite{ken95,damour_vokr}.

\section{New LLR Tests of Relativity}
\label{sec:EP}

Each observation used in this analysis is a measured round-trip light time, here called a ``range,'' between an observatory and a retroreflector. For data processing, the ranges represented by the returned photons are statistically combined into a normal point; each normal point comprising from 3 up to $\sim$100 photons. This paper's analysis of the LLR data from March 1970 to April 2004 uses a total of 15,553 LLR normal points from McDonald and Haleakala Observatories and Observatoire de la C\^ote d'Azur. For the last 10 years of ranges the weighted rms scatter after the fits is $\sim2$ cm. This scatter is $0.5 \times 10^{-10}$ relative to the 385,000 km mean distance of the Moon. 
 
All fits of the lunar laser ranges involve a number of standard solution parameters for the Earth, Moon and lunar orbit (see \cite{[62],oxnard} for the relativistic model used for JPL solutions). The ephemerides for the Moon and planets plus the rotation of the Moon are generated by a simultaneous numerical integration. Least-squares solutions require partial derivatives of range with respect to all solution parameters. Partial derivatives for the lunar orbit and rotation variations with respect to solution parameters are generated by numerical integration.

\subsection{ Equivalence Principle Solution}

In essence, LLR tests of the EP compare the free-fall accelerations of the Earth and Moon toward the Sun.  If the EP is violated, the lunar orbit will be displaced along the Earth-Sun line producing a range signature having a 29.53 day synodic period \cite{[39],ken95,[22aw]} (different from the lunar orbit period of 27 days). Since the first LLR tests of the EP were published in 1976 \cite{[17]}, the precision of the test has increased by two orders-of-magnitude \cite{[62],[2]}. 

The EP test is sensitive to the difference in $M_G/M_I$ between the Earth and Moon. 
A test of the EP, corrected for solar radiation pressure \cite{vokr}, is obtained from a fit of LLR data 	
\begin{equation}
\left[\left(\frac{M_G}{M_I}\right)_e-
\left(\frac{M_G}{M_I}\right)_m\right]_{EP} = (-1.0\pm1.4)\times10^{-13}.
\label{eq:4spc}
\end{equation}
This is equivalent to an orbit perturbation of $\Delta r=(2.8\pm4.1)$ mm $\cos D,$ where angle $D$ corresponds to the 29.53~day mean period of the new--full--new Moon cycle.

\subsection{Equivalence Principle Implications}

The LLR result (\ref{eq:4spc}) is a strong test of the EP. This test is sensitive to  violations due to composition and gravitational self-energy. A University of Washington (UW)  laboratory EP experiment \cite{[1]} is designed to simulate the compositional differences of the Earth and Moon. That test of the relative acceleration is $(1.0\pm1.4)\times 10^{-13}$, where systematic and random uncertainties  are combined \cite{[1]}. The laboratory results are insensitive to self-energy. A combination of
the UW composition test with the LLR result (Eq.~\ref{eq:4spc}) yields the following result for the SEP test
{}
\begin{equation}
\left[\left(\frac{M_G}{M_I}\right)_e-
\left(\frac{M_G}{M_I}\right)_m\right]_{SEP}=(-2.0\pm2.0)\times10^{-13}.
\label{eq:(5n)}
\end{equation} 

Tests for violations of the EP due to self-energy are sensitive to a linear combination of the PPN quantities, Eq.~(\ref{3}). Considering only PPN $\beta$ and $\gamma$, combine Eqs.~(\ref{eq:sep_llr}) and (\ref{eq:(5n)}) to obtain 
\begin{equation}
\eta = 4\beta - \gamma-3 = (4.4\pm4.5)\times 10^{-4}.
\label{eq:(6)}
\end{equation}
This expression would be null for general relativity, hence the small value is consistent with Einstein's theory.

The SEP relates to the non-linearity of gravity (how gravity affects itself), with the PPN parameter  $\beta$ representing the degree of non-linearity. Thus, LLR provides great sensitivity to $\beta$, as suggested by the strong dependence of $\eta$ on  $\beta$  in Eqs.~(\ref{3}) and (\ref{eq:(6)}). The parameter $\gamma$  has been measured independently via time-delay and gravitational ray-bending techniques. The published Viking \cite{[47]} and Very Long-Baseline Interferometry (VLBI) \cite{[48]} uncertainties for $\gamma$  of $\sim$0.002 led to a $\beta$ uncertainty of $\sim$0.0005 dominated by the uncertainty in $\gamma$ \cite{[2]}. 

A much more accurate result for $\gamma$ was recently reported by the Cassini experiment; the test provided a verification that $\gamma$ is unity to a very high accuracy  $\gamma-1=(2.1\pm2.3)\times 10^{-5}$  \cite{cassini_ber}. This leads to a significant improvement in the parameter $\beta$ derived from $\eta$. Combining the Cassini spacecraft determination of $\gamma$ with the $\eta$ of Eq.~(\ref{eq:(6)}), determined from the LLR test of the EP and laboratory WEP results, gives
\begin{equation}
\beta-1=(1.2\pm1.1)\times 10^{-4}.
\label{eq:(7)}
\end{equation}
This result is not a significant deviation of $\beta$ from unity. 

\subsection{LLR Tests of Other Gravitational Physics Parameters}

In addition to the SEP constraint Eq.~(\ref{eq:(6)}), the PPN parameters  $\gamma$ and $\beta$  affect the orbits of relativistic point masses, and $\gamma$  also influences time delay \cite{[62]}. LLR can test this orbital $\beta$  and  $\gamma$ dependence, as well as geodetic de-Sitter precession, and $\dot G/G$ \cite{[62],science94,jim01}.  The possibility of a time variation of the constant of gravitation, {\it G}, was first considered by Dirac in 1938 on the basis of his large number hypothesis, and later developed by Brans and Dicke  in their theory of gravitation (for more details consult \cite{[58]}). Variation might be related to the expansion of the Universe, in which case $\dot G/G=\sigma H_0$, where $H_0$ is the Hubble constant and $\sigma$ is a dimensionless parameter whose value depends on both the gravitational constant and the cosmological model considered. Revival of interest in the Brans-Dicke-like theories, with a variable {\it G}, is partially motivated by the appearance of superstring theories where {\it G} is a dynamical quantity \cite{Marciano1984}. 

In this LLR analysis, the test of temporal variation of the gravitational constant results in  
\begin{equation}
\dot{G}/{G} = (4\pm 9)\times10^{-13} ~~{\rm yr}^{-1}
\end{equation}
with a largest correlation of 0.74 with the diurnal tidal dissipation parameter. The $\dot{G}/G$ uncertainty is 83 times smaller than the inverse age of the Universe, $t_0=13.4$~Gyr with the value for the Hubble constant $H_0=72$~km/sec/Mpc from the WMAP data \cite{Spergel:2003cb}. Any isotropic expansion of the Earth's orbit which conserves angular momentum will mimic the effect of $\dot{G}$ on the Earth's semimajor axis,  $\dot{a}/a=-\dot{G}/G$ \cite{[62]}. There is no evidence for such local ($\sim$1 AU) scale expansion of the solar system.  The uncertainty for $\dot G/G$ is improving rapidly because the sensitivity for new observations depends on the square of the time span. 

The test of geodetic precession yields 
\begin{equation}
K_{gp} = -0.0019 \pm 0.0064. 
\end{equation}
$K_{gp}$ is a relative deviation of geodetic precession from its GR value. The geodetic precession is highly correlated (0.88) with the lunar potential Love number and a parameter for lunar core oblateness.  Adding the latter parameter, not present in the earlier solutions \cite{[62],science94,jim01}, increases the uncertainty of the geodetic precession. 
 
\section{Conclusions} 
\label{sec:conc}

The LLR data set provides sensitive tests of EP, PPN $\beta$, geodetic precession, and $\dot{G}/G$. There have been major improvements in the solution uncertainties since the 1996 results \cite{[62]}.  This improvement is partly due to an additional decade of high quality ranges and partly due to improvements in the model and data fits.  The matter of energy dissipation in the Moon, which previously was a limitation for $\dot{G}/G$ in \cite{[62]}, is now much better understood \cite{williams2001}.  For geodetic precession, the influences on classical precession rates due to inclination and lunar $J_2$ are much improved because of the added data span, but lunar core oblateness, a recent addition to the set of solution parameters, is found to influence classical precession rates through its correlation with tidal distortion.  The LLR EP test has improved markedly in the past decade, but much of that improvement was present in \cite{[2]}.  Since that paper, the laboratory results for the WEP have improved by a factor of two \cite{[1]} and the determination of $\gamma$ has improved by two orders-of-magnitude \cite{cassini_ber} yielding a five-fold improvement in PPN $\beta$.  Increased data span and future improved accuracy should continue to improve LLR tests of gravitational physics.  The high accuracy LLR station being installed at Apache Point \cite{[35]} should provide major opportunities.

\subsubsection*{Acknowledgments~~} 
We acknowledge contributions of the laser ranging stations at McDonald and Haleakala Observatories, and the Observatoire de la C\^ote d'Azur.  The work described here was carried out at the Jet Propulsion Laboratory, California Institute of Technology, under a contract with the National Aeronautics and Space Administration.



\end{document}